\documentclass[11pt,twoside,reqno]{amsart}
\usepackage{a4wide}
\usepackage{amsmath,amssymb} 
\usepackage{graphicx} 
\usepackage[utf8]{inputenc} 
\usepackage[font=small,labelfont=bf]{caption}
\usepackage{bbm}

\begin{document}

\newtheorem{thm}{Theorem}[section]
\newtheorem{lem}[thm]{Lemma}
\newtheorem{prop}[thm]{Proposition}
\newtheorem{coro}[thm]{Corollary}
\newtheorem{defn}[thm]{Definition}
\newtheorem*{remark}{Remark}

\numberwithin{equation}{section}

\newcommand{\Z}{{\mathbb Z}} 
\newcommand{\Q}{{\mathbb Q}}
\newcommand{\PP}{{\mathbb P}}
\newcommand{\R}{{\mathbb R}}
\newcommand{\C}{{\mathbb C}}
\newcommand{\N}{{\mathbb N}}
\newcommand{\FF}{{\mathbb F}}
\newcommand{\T}{{\mathbb T}}
\newcommand{\fq}{\mathbb{F}_q}
\newcommand{\IS}{\mathbb{S}}

\newcommand{\fixmehidden}[1]{}

\def\scrA{{\mathcal A}}
\def\cB{{\mathcal B}}
\def\cC{{\mathcal C}}
\def\Eps{{\mathcal E}}
\def\cI{{\mathcal I}}
\def\scrD{{\mathcal D}}
\def\cF{{\mathcal F}}
\def\cL{{\mathcal L}}
\def\cM{{\mathcal M}}
\def\cN{{\mathcal N}}
\def\cP{{\mathcal P}}
\def\scrR{{\mathcal R}}
\def\scrS{{\mathcal S}}
\def\j{\mathfrak{j}}

\newcommand{\rmk}[1]{\footnote{{\bf Comment:} #1}}

\renewcommand{\mod}{\;\operatorname{mod}}
\newcommand{\ord}{\operatorname{ord}}
\newcommand{\TT}{\mathbb{T}}
\renewcommand{\i}{{\mathrm{i}}}
\renewcommand{\d}{{\mathrm{d}}}
\renewcommand{\^}{\widehat}
\newcommand{\HH}{\mathbb H}
\newcommand{\Vol}{\operatorname{vol}}
\newcommand{\area}{\operatorname{area}}
\newcommand{\tr}{\operatorname{tr}}
\newcommand{\norm}{\mathcal N} 
\newcommand{\intinf}{\int_{-\infty}^\infty}
\newcommand{\ave}[1]{\left\langle#1\right\rangle} 
\newcommand{\E}{\mathbb E}
\newcommand{\Var}{\operatorname{Var}}
\newcommand{\Cov}{\operatorname{Cov}}
\newcommand{\Prob}{\operatorname{Prob}}
\newcommand{\sym}{\operatorname{Sym}}
\newcommand{\disc}{\operatorname{disc}}
\newcommand{\CA}{{\mathcal C}_A}
\newcommand{\cond}{\operatorname{cond}} 
\newcommand{\lcm}{\operatorname{lcm}}
\newcommand{\Kl}{\operatorname{Kl}} 
\newcommand{\leg}[2]{\left( \frac{#1}{#2} \right)}  
\newcommand{\id}{\operatorname{id}}
\newcommand{\beq}{\begin{equation}}
\newcommand{\eeq}{\end{equation}}
\newcommand{\bsp}{\begin{split}}
\newcommand{\esp}{\end{split}}
\newcommand{\bra}{\left\langle}
\newcommand{\ket}{\right\rangle}
\newcommand{\diam}{\operatorname{diam}}
\newcommand{\supp}{\operatorname{supp}}
\newcommand{\dist}{\operatorname{dist}}
\newcommand{\sgn}{\operatorname{sgn}}
\newcommand{\inte}{\operatorname{int}}
\newcommand{\re}{\operatorname{Re}}
\newcommand{\mubar}{\overline{\mu}}

\newcommand{\sumstar}{\sideset \and^{*} \to \sum}

\newcommand{\LL}{\mathcal L} 
\newcommand{\sumf}{\sum^\flat}
\newcommand{\Hgev}{\mathcal H_{2g+2,q}}
\newcommand{\USp}{\operatorname{USp}}
\newcommand{\conv}{*}
\newcommand{\CF}{c_0} 
\newcommand{\kerp}{\mathcal K}

\newcommand{\gp}{\operatorname{gp}}
\newcommand{\Area}{\operatorname{Area}}

\title [Multifractal eigenfunctions in the thermodynamic limit]{Multifractal scaling of quantum eigenfunctions in the thermodynamic limit}
  
\author{Jonathan P. Keating} 
\address{Mathematical Institute, University of Oxford, Andrew Wiles Building, Radcliffe Observatory Quarter, Woodstock Road, 
Oxford OX2 6GG, UK}
\email{jon.keating@maths.ox.ac.uk}

\author{Henrik Uebersch\"ar}
\address{Sorbonne Universit\'e and Universit\'e Paris Cit\'e, CNRS, IMJ-PRG, F-75005 Paris, France.}
\email{henrik.ueberschar@imj-prg.fr}
\date{\today}

\date{\today}
\maketitle

\begin{abstract}
We prove multifractal scaling of measures related to the eigenfunctions of Schr\"odinger operators with point-like bump potentials, in the thermodynamic limit, in dimension $d\geq 5$.  This is consistent with a longstanding expectation in the physics literature in these dimensions. The fractal exponents are computed explicitly as a function of the parameter defining the moments.  Our result requires a weak disorder condition on the location of the scatterers and a strong coupling condition on the sequence of eigenvalues.
\end{abstract}

\section{Introduction}
\subsection{Historical background}
One of the most important open problems in mathematical physics is to give a rigorous proof of the localization-delocalization transition in the Anderson model for electron transport in a disordered medium. Ever since Anderson discovered in 1958 that the presence of disorder may prevent electron transport \cite{PWA}, an extensive physics and mathematical literature has developed around this phenomenon. The Anderson transition has been well-understood  by physicists since the 1970s thanks to a scaling theory \cite{AALR79} which predicts its existence in dimension $d\geq3$ if the disorder is sufficiently weak. The progress of mathematicians, however, has been much slower. 

In 1983, Fr\"ohlich and Spencer \cite{FS83} proved Anderson localization for sufficiently strong disorder in any dimension. Molchanov and Aizenman \cite{AM93} gave an alternative proof in the 1990s. Although a rigorous understanding of the localized regime has been achieved in a wide variety of models over the past few decades \cite{DaSi02,BoKe05,GeKl13,KLNS12}, the delocalized regime, which is conjectures to exist at weak disorder, and the phase transition separating the localized and delocalized regimes, have continued to defy all efforts by mathematicians until now. Remarkable progress has been made in recent years for a toy model, so-called {\em random band matrices}, where both the localized and delocalized regimes are now understood at the rigorous level \cite{B19, Y26}. Moreover, in their release of over 700 mathematical manuscripts on Oct. 6, 2026, Open AI claimed a proof of spectral Anderson delocalization in the lattice model \cite{OAI26} for sufficiently weak disorder in dimensions greater or equal than three.

In 1980, Wegner made the important suggestion \cite{W80} that the Anderson transition should be characterised by the anomalous scaling of the moments of the eigenfunctions with respect to system size in the thermodynamic limit. Such {\em multifractal scaling} has been at the heart of physicists' understanding of the transition between the localizaton and delocalization \cite{ME00,SchrGr91} and is closely related to the appearance of spectral statistics which are {\em intermediate} between Poisson and random-matrix statistics. A mathematical proof of multifractal scaling of eigenfunctions has so far only been obtained in simple toy models where explicit formulae for the eigenfunctions were available \cite{KU21}.

\subsection{Mathematical background}

In this article, we consider the sequence of operators 
\beq\label{gen}
H_L=-\Delta+ \sum_{\omega\in\Omega_L}\alpha_{\omega,L}V_r(\cdot-\omega), \quad a_{\omega,L}\in\C,
\eeq
on the tori $\T^d_L=\R^d/2\pi L\Z^d$, $d\geq5$, as $L\to+\infty$, where $\tilde{V}\in C^\infty_c(\R^d,\C)$, $\supp \tilde{V}\subset B(0,1)$, $\int \tilde V=1$, is a smooth bump potential and
$$V_r(x)=\sum_{k\in 2\pi\Z^d}\tilde{V}_r(x-Lk), \quad \tilde{V}_r(x)=r^{-d}\tilde{V}\left(\frac{x}{r}\right),$$
is the periodized potential which lives on $\T^d_L$. 

Throughout this paper we make the following assumptions:

\begin{itemize}
\item[(i)]{The set $\Omega_L$ is the finite point set encoding the {\em locations of the scatterers} on $\T^d_L$. 
We refer to the {\em thermodynamic limit}, if $\#\Omega_L\sim\rho L^d$, for $\rho>0$, as $L\to+\infty$, and we consider eigenvalues lying in an energy window $[E_1,E_2]$ for fixed $150\leq E_1<E_2$.} \\

\item[(ii)]{We consider a  strictly increasing sequence of length scales $(L_j)_{j=1}^{+\infty}$, $L_j\nearrow+\infty$, and a sequence of eigenfunctions $u_j$ of the operators $H_{L_j}$, $H_{L_j} u_j=\lambda_j u_j$, such that the eigenvalues $\lambda_j$ converge to a limit $E\in[E_1,E_2]$, as $L_j\to+\infty$.} \\

\item[(iii)]{We let $r=r_j\searrow 0$, as $j\to+\infty$, meaning that the bump potentials used to model the scatterers are {\em point-like}.  This is an {\em essential simplification} which our proof rests upon.}\\
\end{itemize}

{\em Multifractal scaling for probability measures:}

To put Wegner's characterisation of the Anderson transition in terms of the multifractal scaling of eigenfunctions in a mathematical context, let us consider our sequence of eigenfunctions $u_j$ on the sequence of tori $\T^d_{L_j}$, where $(L_j)$ is our sequence of length scales which tends to infinity.  Denote the normalised square amplitude of the eigenfunctions by $$a_j(x)=|u_j(x)|^2/\|u_j\|^2_{L^2(\T^d_{L_j})}$$. 
Let us consider the sequence of super-level sets
$$S_j^\epsilon=\{x\in \T^d_{L_j} \mid a_j(x)\geq (1-\epsilon)\|a_j\|_{L^\infty(\T^d_{L_j})}\}, $$
for some fixed  $\epsilon\in(0,1)$.

Wegner suggested that in the thermodynamic limit the scaling behaviour of the sequence of super-level sets $(S_j^\epsilon)$ characterises the Anderson transition.  In the localised regime, where the amplitudes $(a_j)$ remain concentrated in a small region, as $L_j\to+\infty$, one observes few isolated islands in the region where mass concentrates. In the delocalised regime one observes many islands of similar size whose number grows like the volume of the torus. At critical disorder, however, one observes a much more complex behaviour. As the volume tends to infinity, more and more smaller islands appear to form a highly complex fractal pattern. This self-similarity cannot be captured by a single or even a finite number of scaling exponents, but rather by a continuum of exponents. As a mathematical criterion to capture this behaviour Wegner suggested to compute the scaling exponents for the moments of the amplitudes $(a_j)$.

Let us define what we mean by multifractal scaling for a sequence of probability measures. Scaling laws and fractal exponents serve as a tool in the field of multifractal analysis, which studies the self-similar features of sequences of probability measures. Such sequences can display a monofractal self-similarity, or even a much more complex type of self-similarity which we will refer to as multifractal. A standard tool which allows to distinguish these features is the fractal spectrum of such sequences which can be computed via their Renyi entropy.

\begin{defn}
Let $\mu_N$ be a  sequence of probability measures with a sample space $X_N$ of cardinality $|X_N|=N$. We say that the sequence $(\mu_N,X_N)$ satisfies a multifractal scaling law if $\mu_N$ satisfies a power law for $q\in[q_1,q_2]$, $q_2>q_1>1$,
\beq
\sum_{\omega\in X_N}\mu_N(\omega)^q=N^{(1-q)D_q+o(1)}, \quad \text{as}\; N\to+\infty,
\eeq
and the set $\Sigma=\{D_q \mid q\in[q_1,q_2]\}$ contains a non-empty open interval. We refer to $N$ as the {\bf scaling parameter} and to $\Sigma$ as the {\bf fractal spectrum} associated with the sequence $(\mu_N,X_N)$.

In particular, we define the fractal exponent as
\beq
D_q=\lim_{N\to+\infty}\frac{H_q(\mu_N)}{\log N}, \quad q>1,
\eeq
where
\beq
H_q(\mu_N)=\frac{1}{1-q}\log(\sum_{\omega\in X_N}\mu_N(\omega)^q)
\eeq
denotes the {\bf Renyi entropy} of the measure $\mu_N$. 

If a measure is not a probability measure, then we define its Renyi entropy as the Renyi entropy of the associated probability measure obtained by normalisation on $X_N$.
\end{defn}

We note that a sequence $(\mu_N,X_N)$ exhibits monofractal self-similarity if the fractal spectrum contains only a single element. In the case of monofractal scaling, a single global scaling exponent is sufficient to describe the self-similarity of $(\mu_N)$. In the case of multifractal scaling, a continuous spectrum of scaling exponents is required. This means that the scaling exponent is local, because it depends on the location at which the scaling occurs. The parameter $q>1$ amplifies regions, where the measure is large and the computation of $D_q$ for a range of parameters $q$ allows us to distinguish multifractal from monofractal self-similarity.\\

{\em The semi-classical regime:}

For the moment, before we address the case of the thermodynamic limit, let us consider a {\em fixed} general domain $D\subset\R^d$. To understand how the abstract formalism of multifractal scaling can be applied to a sequence of Laplace eigenfunctions $u_j$, as $\lambda_j\to+\infty$, let us associate with each $L^2$-normalised eigenfunction $u_j$ the probability measure induced by the density $d\mu_j(x)=|u_j(x)|^2 dx$ on $D$. 

Let us decompose $D$ into a union of hypercubes of side-length $\epsilon$:
\begin{equation}
D=\bigcup_{\xi\in\Z^d: \epsilon B_\xi\cap D\neq \emptyset} \epsilon B_\xi \cap D
\end{equation}
where $B_\xi=[\xi_1,\xi_1+1)\times\cdots\times[\xi_d,\xi_d+1)$.

In this way we obtain a sequence of discrete probability measures which arise from the probabilities that each hypercube is charged with by the density $d\mu_j$: $p(\xi)=\int_{\epsilon B_\xi\cap D}d\mu_j$. The scaling parameter  is given by the number of hypercubes $N_\epsilon=\#\{\xi\in\Z^d \mid \epsilon B_\xi\cap D\neq\emptyset\}\sim \epsilon^{-d}\Vol(D)$, as $\epsilon\to0$. We can think of $\epsilon$ as the scale at which we ``zoom in'' to the structure of the measure $d\mu_j$, and we cannot shrink faster than the Planck scale:\footnote{We use the notation $f\gtrsim g$ to denote that there exists a constant $C>0$ s. t. $f\geq Cg$.} $\epsilon\gtrsim \lambda_j^{-1/2}$, since the eigenfunctions do not oscillate below this scale.\\

{\em The thermodynamic regime:}

In the case of the thermodynamic limit, we {\em fix a spectral window}, and, therefore, the wavelength. On the other hand, the volume of the domain tends to infinity: $\T^d_L$, $L\to+\infty$. In this regime we apply a decomposition of unity with hypercubes of fixed side length. Hence, the scaling parameter $N=N_L$ tends to infinity, as $L\to+\infty$.

An alternative approach, which is particularly powerful in the case of toral Schr\"odinger operators, was presented in \cite{KU21}. In this case, we studied the eigenfunctions in their momentum representation which gives rise to a sequence of probability measures on the Fourier lattice $\Z^d$. The bulk of the mass of the measure concentrates on an annulus with central radius $\sqrt{\lambda_j}$. In the thermodynamic limit, however, the wavelength remains bounded and the Fourier lattice scales with $L$: $L^{-1}\Z^d$. Hence, we are dealing with a measure whose bulk is concentrated in a fixed ball.\\

\subsection{Definition of the averaged measure on $L_j^{-1}\Z^d$}

Our ultimate goal is to prove multifractal scaling laws for measures associated with the eigenfunctions of a sequence of toral Schr\"odinger operators. We will follow the approach outlined in \cite{KU21} by studying probability measures associated with the eigenfunctions on the Fourier side. As in \cite{KU21}, a key challenge is the determination of the scaling parameter by restricting the measure on the infinite lattice $L_j^{-1}\Z^d$ to a finite ball. Due to the decay of the Fourier coefficients of the eigenfunctions, the probability measure on the infinite lattice will be dominated by a finite number of contributions. The scaling parameter is the number of these contributions and, as we will see, it will depend on the parameters $r_j$ and $L_j$, as well as the dimension $d$.

Throughout this article we will define the Fourier transform of $u_j$ as follows:
\footnote{We intentionally omit the normalization factor $(2\pi L_j)^{-d/2}$ in this definition.}
$$\hat{u}_j\left(\frac{\xi}{L_j}\right):=\int_{\T^d_{L_j}}u_j(x)e^{-i \frac{\xi \cdot x}{L_j}}dx.$$

Before we define the measure which is the object of study in this article, we introduce the following measure on the lattice $L_j^{-1}\Z^d$:
$$\nu_j\left(\frac{\xi}{L_j}\right):=\frac{\Big|\widehat{u_j}\left(\frac{\xi}{L_j}\right)\Big|^2}{\|u_j\|_{L^2(\T^d_{L_j})}^2}$$

By applying the Fourier transform to the eigenvalue problem $H_{L_j}u_j=\lambda_j u_j$, which we rewrite as $(-\Delta-\lambda_j)u_j=-V_{L_j}u_j$, we obtain
\beq
|\widehat{u_j}(\frac{\xi}{L_j})|^2=L_j^4\frac{|\widehat{V_{L_j}u_j}(\xi/L_j)|^2}{||\xi|^2-\lambda_j L_j^2|^2}
\eeq
where we denote $$V_{L}=\sum_{\omega\in\Omega_L}\alpha_{\omega,L}V_r(\cdot-\omega).$$

{\em We recall our assumption $(iii)$: $r_j\searrow 0$, as $L_j\to+\infty$.} This means the potentials $V_{r_j}$ are ``point-like'', by which we mean that the diameter of the support of $V_{r_j}$ shrinks to zero in the thermodynamic limit. 

We expand as a sum over Fourier integrals at the scatterer locations
\beq\label{linear-split}
\widehat{V_{L_j}u_j}(\xi/L_j)=\sum_{\omega\in\Omega_{L_j}}\alpha_{\omega,L_j}\int_{\T^d_{L_j}}V_{r_j}(x-\omega)u_j(x)e^{-i\tfrac{\xi}{L_j}\cdot x}dx
\eeq
and at each location $\omega\in\Omega_{L_j}$ we have for any Fourier frequency which corresponds to a {\em wavelength which is large compared with the diameter $r_j$ of the bump potential}: $L_j/|\xi|\geq C r_j$, for some large constant $C>0$,
\beq\label{s-wave}
\int_{\T^d_{L_j}}V_{r_j}(x-\omega)u_j(x)e^{-i\tfrac{\xi}{L_j}\cdot x}dx=u_j(\omega)e^{-i\tfrac{\xi}{L_j}\cdot \omega}+O_{u_j}\left(\frac{|\xi|r_j}{L_j}\right)
\eeq
where we refer to the first term as the {\em s-wave component}.

On the other hand, Fourier coefficients of a generic potential decay rapidly in a regime, where the wavelength is much smaller than the radius of the ball on which $V_{r_j}$ is supported: $L_j/|\xi|\ll r_j$. In fact, repeated integration by parts gives the bound: 
\beq
\left|\int_{\T^d_{L_j}}V_{r_j}(x-\omega)u_j(x)e^{-i\tfrac{\xi}{L_j}\cdot x}dx\right|=O_{u_j,V,\kappa}\left(\left(\frac{L_j}{|\xi|r_j}\right)^\kappa\right).
\eeq

We will study the multifractal scaling properties of the discrete measure on $L_j^{-1}\Z^d$ obtained by replacing each Fourier integral in \eqref{linear-split} by the s-wave component in \eqref{s-wave} at the location $\omega\in\Omega_{L_j}$ for all frequencies which satisfy $|\xi|/L_j\leq C^{-1}r_j^{-1}$. 
\begin{defn}[s-wave component of $\nu_j$]
We define the measure
\beq
\mu_j\left(\frac{\xi}{L_j}\right):=\frac{\Big|\sum_{\omega\in\Omega_{L_j}}\alpha_{\omega,L_j} u_j(\omega)e^{-i \xi\cdot\omega/L}\Big|^2}{||\xi|^2-\lambda_j L_j^2|^2}
\eeq
on the sample space of lattice vectors inside a ball of radius $R_j=1/Cr_j$
\beq
X_j=L_j^{-1}\Z^d\cap B(0,R_j).
\eeq
\end{defn}

Note that, for simplicity, we drop the factor $L_j^4/\|u_j\|^2_{L^2(\T^d_{L_j})}$ in the definition. This is possible, because the normalisation of the measure in the definition of the Renyi entropy ensures that any factor cancels out.

The scaling parameter is given by $N_j=|X_j|\sim c_d (R_jL_j)^d$. We note that the choice of the constant $C>0$ plays no role in the computation of the fractal exponents $D_q$, because its choice does not affect the leading term in the asymptotic of $\log N_j$.

\subsection{Main result}
Let $h\in\C^\infty_c(\R^d)$ be positive with $\supp h\subset B(0,1)$ and $\int_{\R^d}h=1$. Fix $N>d$. We have $|\hat{h}(k)|\leq C_N|k|^{-N}$ for some constant $C_N>0$. 

For a given function $f:L^{-1}\Z^d \to\C$ we define
\beq
\langle f\rangle_h := L^{-d}\sum_{\zeta\in L^{-1}\Z^d} h(\zeta)f(\zeta)
\eeq

We define a new measure $\mubar_j$ by averaging $\mu_j$ over $B(k,1)\cap L_j^{-1}\Z^d$
\beq
\mubar_j(k):=\langle f\rangle_{h(\cdot-k)} 
\eeq
for any $k\in L_j^{-1}\Z^d$.

We make the following assumptions:
\begin{itemize}
\item[\bf (WD)] Assume that we satisfy a {\bf weak disorder} condition in the sense that for any $\omega\in\Omega_{L_j}$ we have
\beq
\sum_{\omega'\in\Omega_{L_j}\setminus\{\omega\}}|\omega-\omega'|^{-N} \leq \frac{1}{2C_N}.
\eeq
\\

\item[\bf (SC)] Moreover, assume that the sequence of eigenvalues $\lambda_j$ satisfies a {\bf strong coupling} condition: 
\beq
|\lambda_j L_j^2-n_j^\pm|\gtrsim 1,
\quad \text{as} \; j\to+\infty,
\eeq
where $n_j^{\pm}$ denote the neighbouring integers of $\lambda_jL_j^2$.
\\
\end{itemize}

We first have the following result which confirms that there exists a sequence of operators which satisfies these conditions.
\begin{thm}\label{thm-exist}
There exists a sequence of length scales $(L_j)$ and operators $H_{L_j}$ such that {\bf (WD)} and {\bf (SC)} are satisfied.
\end{thm}

We may compute the fractal exponents of the sequence of measures $\mubar_j$ in dimension $d\geq 5$ in a point-like regime provided the radius $r_j$ decays sufficiently rapidly, as $j\to+\infty$.
\begin{thm}\label{thm-mult}
Let $d\geq 5$. Assume that $r_j\searrow 0$ sufficiently fast s. t. $\log L_j=o(\log r_j^{-1})$, as $j\to+\infty$. Then, under assumptions {\bf (WD)} and {\bf (SC)}, we have that $D_q=\frac{d-4}{d}\frac{q}{q-1}$ for any $q>d/4$.
\end{thm}

The fact that $D_q$ is not a constant function of $q$ means that the measures $\mubar_j$ exhibit multifractal scaling when $d\geq 5$, in the sense defined above. \\

{\bf Remark: Dimensions three and four.} Our proof makes crucial use of the assumption $d\geq 5$. While multifractal scaling is predicted in dimension $d\geq 3$ in the physics literature, and
the formula we find for $D_q$ vanishes if we formally set $d=4$, it is important to point out that this allows {\em no conclusion} to be drawn regarding the form of $D_q$ in this case. We prove an asymptotic for the Renyi entropy which is based on asymptotic upper and lower bounds for the number of lattice points on a sphere of radius $\sqrt{n}$ in dimension $d$, the arithmetic function $r_d(n)$. The bounds we need are not available in dimensions $d=3,4$. 

Moreover, our ability to obtain an asymptotic for the Renyi entropy, as $L_j\to+\infty$, hinges on the fact that a leading growing term dominates another more complicated, potentially oscillatory term, which is only the case in dimension $d\geq 5$. At present, we are unable to adapt our methods for dimensions $d=3,4$.  It may be that an entirely different approach is necessary to prove multifractality in these cases.

\subsection{Outline of the proof}

We briefly describe the main idea of our proof. Let us denote the $q$th moment of the averaged measure $\mubar_j$:
\beq
M_{q,j}=\sum_{k\in L_j^{-1}\Z^d\cap B(0,R_j)}\mubar_j(k)^q.
\eeq
The Renyi entropy of $\mubar_j$ is defined as 
\beq
\begin{split}
H_q(\mubar_j)&=\frac{1}{1-q} \log \left(\sum_{k\in L_j^{-1}\Z^d\cap B(0,R_j)} \left(\frac{\mubar_j(k)}{M_{1,j}}\right)^q\right)\\
&=\frac{\log M_{q,j}-q\log M_{1,j}}{1-q}
\end{split}
\eeq

Let $$f(L_j):=\sum_{\omega\in\Omega_L}|\alpha_{\omega,L_j}u_j(\omega)|^2.$$
Under assumptions {\bf (WD)} and {\bf (SC)} we prove that the rescaled moments $M'_{q,j}=f(L_j)^{-q}M_{q,j}$ satisfy
$$\log M'_{1,j}=(d-4)\log R_j+d\log L_j+O(1)$$
and
$$d\log L_j \leq \log M'_{q,j} \leq (4q+d)\log L_j$$
as $L_j\to+\infty$.

From this we may then compute the asymptotic of $H_q(\mubar_j)$, which allows us to compute the fractal exponents $D_q$ for any $q>d/4$, provided that $r_j\searrow0$ fast enough, as $j\to+\infty$. \\

{\bf Organisation of the paper:}
The paper is structured as follows. The proof of multifractal scaling in dimension $d\geq 5$, Theorem \ref{thm-mult}, is presented in section 2. In section 3 we give the proof of Theorem \ref {thm-exist}.  While the methods introduced in this paper do not suffice to prove multifractality in dimension $d=3,4$, we are able to prove general limit theorems on the Fourier side for eigenfunctions in the thermodynamic limit. These supplementary results are stated in section 4. In section 5, we motivate the different coupling regimes considered in section 4. Section 6 contains the proofs of the limit theorems stated in section 3. In section 7, we prove asymptotics for lattice point sums which are used in section 4.\\

{\bf AI.}  No use of AI was made in this work, or in writing this manuscript.\\

{\bf Open Access.} For the purpose of Open Access, the authors have applied a CC BY public copyright licence to any Author Accepted Manuscript (AAM) version arising from this submission.\\

{\bf Acknowledgements.} The authors would like to thank Jens Marklof for a careful reading of the manuscript and many useful suggestions which led to the improvement of this paper.

\section{Proof of Theorem \ref{thm-mult}}\label{sec-mult}

\subsection{The weak disorder condition}
We start with the following lemma.
\begin{lem}\label{av-lem}
Let $h\in\C^\infty_c(\R^d)$ be positive with $\int_{\R^d}h=1$, and $|\hat{h}(k)|\leq C_N |k|^{-N}$ for some $N>d$ and $C_N>0$.

Assume that for all $\omega'\in\Omega_L$ 
\beq
\sum_{\omega\in\Omega_L\setminus\{\omega'\}}|\omega-\omega'|^{-N} \leq \frac{1}{2 C_N}.
\eeq

For a given function $f:L^{-1}\Z^d \to\C$ we define
\beq
\langle f\rangle_h := L^{-d}\sum_{\zeta\in L^{-1}\Z^d} h(\zeta)f(\zeta).
\eeq

We have
\beq
\langle \Big|\sum_{\omega\in\Omega_{L_j}}\alpha_{\omega,L_j} u_j(\omega)e^{-i \xi\cdot\omega/L_j}\Big|^2 \rangle_h \asymp \sum_{\omega\in\Omega_L} |\alpha_{\omega,L}|^2 |u_j(\omega)|^2
\eeq
for $L_j$ large enough.
\end{lem}
\begin{proof}
We compute
\beq\label{lem-eq1}
\begin{split}
&\langle \Big|\sum_{\omega\in\Omega_{L_j}}\alpha_{\omega,L_j} u_j(\omega)e^{-i \xi\cdot\omega/L_j}\Big|^2 \rangle_h\\
=&L_j^{-d}\sum_{\zeta\in L_j^{-1}\Z^d}h(\zeta)\Big|\sum_{\omega\in\Omega_{L_j}}\alpha_{\omega,L_j} u_j(\omega)e^{-i \zeta\cdot\omega}\Big|^2\\
=& \langle1\rangle_h \sum_{\omega\in\Omega_{L_j}}|\alpha_{\omega,L_j}|^2 |u_j(\omega)|^2\\
&+\sum_{\omega \neq \omega'} \alpha_{\omega,L_j} \overline{\alpha_{\omega',L_j}} u_j(\omega)\overline{u_j(\omega')}\langle e^{i\zeta(\omega'-\omega)}\rangle_h
\end{split}
\eeq

Note that, for $L_j$ sufficiently large,
$$|\langle e^{i\zeta\cdot(\omega'-\omega)} \rangle_h| \leq (1+\epsilon) |\hat{h}(\omega'-\omega)|\leq (1+\epsilon)C_N|\omega-\omega'|^{-N}.$$

Let us denote $\beta_{\omega,L_j}:=\alpha_{\omega,L_j}u_j(\omega)$. 

We estimate 
\beq\label{lem-eq2}
\begin{split}
&\Big|\sum_{\omega\neq \omega'}\beta_{\omega,L_j} \overline{\beta_{\omega',L}} \langle e^{i\zeta(\omega'-\omega)}\rangle_h\Big|\\
&\leq(1+\epsilon)C_N \sum_{\omega\neq\omega'}|\beta_{\omega,L_j}||\beta_{\omega',L_j}| |\omega-\omega'|^{-N}\\
&\leq \frac{1+\epsilon}{2}C_N\sum_{\omega\neq\omega'}(|\beta_{\omega,L_j}|^2+|\beta_{\omega',L_j}|^2)|\omega-\omega'|^{-N}
\end{split}
\eeq

Now, we obtain
\begin{align*}
\sum_{\omega\neq\omega'}|\beta_{\omega,L_j}|^2|\omega-\omega'|^{-N}=&\sum_{\omega\in\Omega_{L_j}}|\beta_{\omega,L_j}|^2\sum_{\omega'\in\Omega_{L_j}\setminus\{\omega\}}|\omega-\omega'|^{-N}\\
\leq& \frac{1}{2 C_N}\sum_{\omega\in\Omega_L}|\beta_{\omega,L_j}|^2
\end{align*}
and, because the second sum in \eqref{lem-eq2} is estimated analogously, the off-diagonal sum in \eqref{lem-eq1} is bounded by 
$\frac{1+\epsilon}{2}\sum_{\omega\in\Omega_{L_j}}|\beta_{\omega,L_j}|^2$.

Hence, we obtain
\beq
\begin{split}
(\langle 1\rangle_h-\tfrac{1+\epsilon}{2})\sum_{\omega\in\Omega_{L_j}}|\beta_{\omega,L_j}|^2\leq &
\left\langle \Big|\sum_{\omega\in\Omega_{L_j}}\alpha_{\omega,L_j} u_j(\omega)e^{-i \xi\cdot\omega/L_j}\Big|^2 \right\rangle_h \\
\leq &(\langle 1\rangle_h+\tfrac{1+\epsilon}{2})\sum_{\omega\in\Omega_{L_j}}|\beta_{\omega,L_j}|^2
\end{split}
\eeq
and the claim follows because $\lim_{L\to+\infty}\langle 1\rangle_h=\int_{\R^d}h=1$.
\end{proof}

\subsection{Computation of the Renyi entropy for the averaged measure}
We consider the averaged measure
\beq
\mubar_j(k):=\sum_{\zeta\in L_j^{-1}\Z^d}h(\zeta-k) \mu_j(\zeta)
\eeq
for any $k\in L_j^{-1}\Z^d$. 

\subsubsection{Estimates on moment sums}
We split the summation into terms near the singularity and its complement
$$
\sum_{k\in  L_j^{-1}\Z^d\cap B(0,R)} \mubar_j(k)^q=\sum_{k\in A}\mubar_j(k)^q+\sum_{k\in B}\mubar_j(k)^q,
$$
where we denote
$$A:=\{k\in  L_j^{-1}\Z^d \cap B(0,R)\mid||k|^2-\lambda_j|<\tfrac{1}{2} |k|^2\}$$
and $$B:=\{k\in  L_j^{-1}\Z^d \cap B(0,R)\mid||k|^2-\lambda_j|\geq \tfrac{1}{2}|k|^2\}$$

First of all, we prove
\begin{itemize}
\item[(a)]{for any $k\in A$ and $\zeta\in\supp h(\cdot-k)\cap L_j^{-1}\Z^d$ we have $L_j^{-2}\lesssim ||\zeta|^2-\lambda_j| \lesssim |\zeta|^{2}$, where we assume that $|\lambda_j L_j^2-n_j^\pm|\gtrsim 1$.}\\
\item[(b)]{for any $k\in B$ and $\zeta\in\supp h(\cdot-k)\cap L_j^{-1}\Z^d$: $||\zeta|^2-\lambda_j| \asymp ||k|^2-\lambda_j|$.}\\
\end{itemize}
\begin{proof}
{\em (a):} First, we have $L_j^2||\zeta|^2-\lambda_j|=||L_j\zeta|^2-\lambda_jL_j^2|\gtrsim 1$, because $L_j\zeta\in\Z^d$ and we recall our assumption $|\lambda_j L_j^2-n_j^\pm|\gtrsim 1$. Hence, we see that $||\zeta|^2-\lambda_j|\gtrsim L_j^{-2}$.

Secondly, $\zeta\in\supp h(\cdot-k)\subset B(k,1)$. 

We denote $\xi=\zeta-k$ and rewrite
\begin{align*}
||\zeta|^2-\lambda_j|&=||k+\xi|^2-\lambda_j|\\
&=||k|^2+2\left\langle\xi,k\right\rangle+|\xi|^2-\lambda_j|.
\end{align*}
Now, we apply the triangle inequality, where we recall that $|\xi|=|\zeta-k|\leq1$, because $\zeta\in \supp h(\cdot-k)\subset B(k,1)$,
\begin{align*}
||k|^2+2\left\langle\xi,k\right\rangle+|\xi|^2-\lambda_j|
&\leq ||k|^2-\lambda_j|+2|k|+1\\
&\leq \frac{1}{2}|k|^2+2|k|+1\\
&\leq \tfrac{3}{4}|k|^2
\end{align*}
where the last inequality holds because  $|k|>10$. To see this, rewrite $$||k|^2-\lambda_j|<\tfrac{1}{2}|k|^2$$ as 
$$-\tfrac{1}{2}|k|^2<|k|^2-\lambda_j<\tfrac{1}{2}|k|^2$$
and, thus, $\lambda_j<\tfrac{3}{2}|k|^2$ and $\tfrac{1}{2}|k|^2<\lambda_j$, which is equivalent to
$$\tfrac{2}{3}\lambda_j < |k|^2 < 2\lambda_j.$$
If we recall $\lambda_j\geq E_1\geq 150$, we see $|k|>\sqrt{\tfrac{2}{3}\lambda_j}\geq 10$ which implies $2|k|+1<\tfrac{1}{4}|k|^2$.

Now $|k|=|\zeta+\xi|\leq |\zeta|+1$, and $\tfrac{3}{4}|k|^2\leq \tfrac{3}{4}(|\zeta|+1)^2<\tfrac{3}{4}(1+\tfrac{1}{9})^2|\zeta|^2=\tfrac{25}{27}|\zeta|^2$, because $|\zeta|\geq |k|-1>9$. Hence $||\zeta|^2-\lambda_j|\lesssim |\zeta|^2$.
\\

{\em (b):} Using the reverse triangle inequality we obtain the lower bound
\begin{align*}
||\zeta|^2-\lambda_j|&=||k+\xi|^2-\lambda_j|\\
&=||k|^2+2\left\langle\xi,k\right\rangle+|\xi|^2-\lambda_j| \\
&\geq ||k|^2-\lambda_j|-2|k|-1
\end{align*}

For $|k|\geq9$ we have
$$ ||k|^2-\lambda_j|-2|k|-1\geq \tfrac{1}{2}||k|^2-\lambda_j|.$$

To see this, we rewrite the inequality above as
$$\tfrac{1}{2}||k|^2-\lambda_j|\geq2|k|+1.$$

Because $k\in B$ and $|k|\geq 9$, we have
$$||k|^2-\lambda_j|\geq \tfrac{1}{2}|k|^2\geq \tfrac{9}{2}|k|>4|k|+2.$$

If $|k|<9$, then $||k|^2-\lambda_j|-2|k|-1\geq ||k|^2-\lambda_j|-39=\lambda_j-|k|^2-39\geq 150-81-39=30$, because $\lambda_j\geq 150$,
and, thus, $$||\zeta|^2-\lambda_j|\geq \tfrac{1}{5}||k|^2-\lambda_j|,$$ in view of $||k|^2-\lambda_j|=\lambda_j-|k|^2\leq 150$.

Hence, we have shown $||\zeta|^2-\lambda_j|\gtrsim ||k|^2-\lambda_j|$.\\

To conclude, let us show $||\zeta|^2-\lambda_j|\lesssim ||k|^2-\lambda_j|$. 

Following the same steps as above,
\begin{align*}
||\zeta|^2-\lambda_j|&=||k+\xi|^2-\lambda_j|\\
&=||k|^2+2\left\langle\xi,k\right\rangle+|\xi|^2-\lambda_j| \\
&\leq ||k|^2-\lambda_j|+2|k|+1\\
&\leq ||k|^2-\lambda_j|+\tfrac{1}{4}|k|^2\\
&\leq \tfrac{3}{2}||k|^2-\lambda_j|.
\end{align*}
where we used in the final inequality that $k\in B$ and, therefore, $\tfrac{1}{2}|k|^2\leq ||k|^2-\lambda_j|$.
\end{proof}

We may now bound the averaged measure from above and below, respectively, in the two ranges:

Define $f(L):=\sum_{\omega\in\Omega_L}|\beta_\omega|^2$.

Recall that 
\beq
\mubar_j(k)=L_j^{-d}\sum_{\zeta\in L_j^{-1}\Z^d}h(\zeta-k)\frac{\Big|\sum_{\omega\in\Omega_{L_j}}\beta_{\omega}e^{-i\zeta\cdot\omega}\Big|^2}{(|\zeta|^2-\lambda_j)^2}
\eeq

If $k\in A$, we employ inequality (a) and Lemma \ref{av-lem} to obtain:
$$\mubar_j(k)\gtrsim L_j^{-d}\sum_{\zeta\in L_j^{-1}\Z^d}h(\zeta-k)|\zeta|^{-4}\Big|\sum_{\omega\in\Omega_{L_j}}\beta_{\omega}e^{-i\zeta\cdot\omega}\Big|^2
\gtrsim |k|^{-4}f(L_j),$$
and
$$\mubar_j(k)\lesssim L_j^{4-d}\sum_{\zeta\in L_j^{-1}\Z^d}h(\zeta-k)\Big|\sum_{\omega\in\Omega_{L_j}}\beta_{\omega}e^{-i\zeta\cdot\omega}\Big|^2\lesssim L_j^4 f(L_j),$$
where we note that for any $k\in L_j^{-1}\Z^d$ we have
$$\left\langle1\right\rangle_{h(\cdot-k)}=L_j^{-d}\sum_{\zeta\in L_j^{-1}\Z^d}h(\zeta-k)=L_j^{-d}\sum_{\zeta\in L_j^{-1}\Z^d}h(\zeta)=\left\langle1\right\rangle_{h}.$$

If $k\in B$, we employ inequality (b) and Lemma \ref{av-lem}:
\begin{align*}
\mubar_j(k) &\asymp (|k|^2-\lambda_j)^{-2}L_j^{-d}\sum_{\zeta\in L_j^{-1}\Z^d}h(\zeta-k)\Big|\sum_{\omega\in\Omega_{L_j}}\beta_{\omega}e^{-i\zeta\cdot\omega}\Big|^2\\
&\asymp (|k|^2-\lambda_j)^{-2}f(L_j)
\end{align*}

Hence, we obtain the following upper and lower bounds for the partial moment sums 
$$f(L_j)^q L_j^d \lesssim \sum_{k\in A}\mubar(k)^q \lesssim L_j^{4q+d} f(L_j)^q,$$
where we used $|A|=\#\{k\in L_j^{-d}\Z^d \mid \tfrac{2}{3}\lambda_j\leq|k|^2\leq 2\lambda_j\}\asymp L_j^d$,
and
$$ \sum_{k\in B}\mubar(k)^q \asymp f(L_j)^q \sum_{k\in B}||k|^2-\lambda_j|^{-2q}.$$

Let us now estimate the sum over terms which are far from the singularity. If $q>d/4$, then we have
\beq
\begin{split}
\sum_{k\in B}||k|^2-\lambda_j|^{-2q}
&=\sum_{\substack{||k|^2-\lambda_j|\geq \tfrac{1}{2}|k|^2 \\ k\in  L_j^{-1}\Z^d\cap B(0,R)}}||k|^2-\lambda_j|^{-2q}\\
&\leq \sum_{0\neq k\in  L_j^{-1}\Z^d\cap B(0,R)}|k|^{-4q}\\
&\leq L_j^{4q}\sum_{n\geq 1}r_d(n)n^{-2q}\lesssim  L_j^{4q}
\end{split}
\eeq
where we define for any $n\in\N$ the number of ways $n$ can be represented as a sum of $d$ squares of integers
\beq
r_d(n)=\#\{\xi\in\Z^d \mid n=|\xi|^2\}
\eeq
and we point out that $r_d(n)\asymp n^{d/2-1}$ for $d\geq5$ (cf. Thm. 20.2, in \cite{IwKo04}, p. 456).

For $q=1$, we have
\beq
\begin{split}
\sum_{k\in B}||k|^2-\lambda_j|^{-2}
&=\sum_{\substack{||k|^2-\lambda_j|\geq \tfrac{1}{2}|k|^2 \\ k\in  L_j^{-1}\Z^d\cap B(0,R)}}||k|^2-\lambda_j|^{-2}\\
&\leq L_j^4\sum_{\substack{||k|^2-\lambda_j L_j^2|\geq c\lambda_j L_j^2 \\ k\in \Z^d\cap B(0,RL_j)}}||k|^2-\lambda_j L_j^2|^{-2}\\
&= \sum_{\substack{|n-\lambda_j L_j^2|\geq c\lambda_jL_j^2 \\ 1\leq n\leq R^2L_j^2}}r_d(n)(n-\lambda_j L_j^2)^{-2}\\
&\lesssim \int_{\substack{|t-\lambda_j L_j^2|\geq c'\lambda_jL_j^2 \\ 1\leq t\leq R^2L_j^2}}t^{d/2-1}(t-\lambda_j L_j^2)^{-2}dt
\end{split}
\eeq
for some constants $c,c'>0$.

Similarly, we can show
\beq
\sum_{k\in B}||k|^2-\lambda_j|^{-2} \gtrsim  \int_{\substack{|t-\lambda_j L_j^2|\geq c''\lambda_jL_j^2 \\ 1\leq t\leq R^2L_j^2}}t^{d/2-1}(t-\lambda_j L_j^2)^{-2}dt
\eeq
for some constant $c''>0$.

Changing variables $s=\lambda_j^{-1}L_j^{-2} t$, we get 
\begin{align*}
&\int_{\substack{|t-\lambda_j L_j^2|\geq c'\lambda_jL_j^2 \\ 1\leq t\leq R^2L_j^2}}t^{d/2-1}(t-\lambda_jL_j^2)^{-2}dt\\
= &\lambda_j^{d/2-2}L_j^{d-4}\int_{\substack{|s-1|\geq c' \\ \lambda_j^{-1}L_j^{-2}\leq s \leq R^2\lambda_j^{-1}}}s^{d/2-1}(s-1)^{-2}ds.
\end{align*}

Let us suppose $R^2\gg \lambda_j$. 

To estimate the integral, we split it into two parts
\begin{align*}
I_1:=&\int_{1+c'}^{R^2\lambda_j^{-1}}s^{d/2-1}(s-1)^{-2}ds\\
&\asymp \int_{1+c'}^{R^2\lambda_j^{-1}}s^{d/2-3}ds\asymp R^{d-4}\lambda_j^{-d/2+2}
\end{align*}
and
$$I_2:=\int_{\lambda_j^{-1}L_j^{-2}}^{1-c'}s^{d/2-1}(s-1)^{-2}ds.$$

We estimate
$$I_2=\int^{1-c'}_{\lambda_j^{-1}L_j^{-2}}s^{d/2-1}(s-1)^{-2}ds \asymp 1,$$
as $j\to+\infty$.

In conclusion, we get $$\sum_{k\in B}||k|^2-\lambda_j|^{-2}\asymp R^{d-4}L_j^d.$$

\subsubsection{Estimate of the Renyi entropy}

We let $R_j=cr_j^{-1}$, where $r_j\to 0$, as $j\to+\infty$.

Define 
\beq
M_{q,j}:=\sum_{k\in\Z^d\cap B(0,R_j)}\mubar_j(k)^q, \quad q\geq 1.
\eeq
Let us introduce the rescaled moments $M_{q,j}':=f(L_j)^{-q}M_{q,j}$.

We have
\beq
H_q(\mubar_j):=\frac{\log M_{q,j}-q\log M_{1,j}}{1-q}=\frac{\log M'_{q,j}-q\log M'_{1,j}}{1-q}.
\eeq

The estimates proved in the previous section,
$$M'_{1,j}\asymp R_j^{d-4}L_j^d$$
and $$L_j^d\lesssim M'_{q,j}\lesssim L_j^{4q+d}, \quad q>d/4$$ yield
$$\log M'_{1,j}=(d-4)\log R_j+d\log L_j+O(1)$$
and
$$d\log L_j \leq \log M'_{q,j} \leq (4q+d)\log L_j.$$

We may now compute the asymptotics for the Renyi entropy
\begin{align*}
\frac{H_q(\mubar_j)}{\log R_j}&=\frac{\log M'_{q,j}-q(d-4)\log R_j+d\log L_j+O(1)}{(1-q)\log R_j}\\
&=\frac{q}{q-1}(d-4)+O_{q,d}\left(\frac{\log L_j}{\log R_j}\right)
\end{align*}
and, if we recall our assumption $\log L_j=o(\log R_j)$, we get
\beq
H_q(\mubar_j)\sim \frac{q}{q-1}(d-4)\log R_j.
\eeq

If we now recall that the scaling parameter is given by $N_j=|\{\xi\in L_j^{-1}\Z^d\cap B(0,R_j)\}|\sim c_d (R_jL_j)^d$, then we get for any $q>d/4$:
\beq
\begin{split}
D_q&=\lim_{j\to+\infty}\frac{H_q(\mu_j)}{\log N_j}\\
&=\lim_{j\to+\infty}\frac{H_q(\mu_j)}{\log R_j}\frac{\log R_j}{\log N_j}\\
&=\frac{q(d-4)}{(q-1)}\frac{1}{d}
\end{split}
\eeq

\section{Proof of Theorem \ref{thm-exist}}

Fix an odd integer $n\geq 1$. Take $L_j=jn$ for odd integers $j\geq 1$. We pick a fundamental domain $[-jn/2,jn/2]^d$ and define
$$\Omega_{L_j}:=\{\tfrac{1}{2}n\xi +\epsilon x_\xi \mid \xi \in \{-j+1,\cdots,j-1\}^d, \;x_\xi\in B(0,1)\}.$$

First, we prove that $\Omega_{L_j}$ satisfies assumption {\bf (WD)}:
Fix $\omega'\in\Omega_{L_j}$. Recall that $N>d$. We will show that, if $n\geq 2\epsilon$, then
\beq
\sum_{\omega\in\Omega_{L_j}\setminus\{\omega'\}}|\omega-\omega'|^{-N}\leq \frac{1}{2C_N}.
\eeq

By translation it is sufficient to consider the case $\omega'=0.$ 
We have for any $\omega\neq 0$: $|\omega|=|n\xi+\epsilon x_\xi|\geq n|\xi|-\epsilon\geq \tfrac{1}{2}n|\xi|$,  where we note that $\omega\neq 0$ implies $|\xi|\geq1$ and $\tfrac{1}{2}n|\xi|\geq \tfrac{1}{2}n\geq\epsilon$.
\begin{align*}
\sum_{\omega\in\Omega_{L_j}\setminus\{0\}}|\omega|^{-N} &\leq \left(\frac{2}{n}\right)^N \sum_{\xi\in\Z^d\setminus\{0\}}|\xi|^{-N}\\
&=\left(\frac{2}{n}\right)^N\sum_{n\geq 1}r_d(n)n^{-N/2}\\
&=\left(\frac{2}{n}\right)^N\zeta_d(N/2)
\end{align*}
and, thus, for
$$n\geq 2(2C_N\zeta_d(N/2))^{1/N}$$
we have
$$\sum_{\omega\in\Omega_{L_j}\setminus\{0\}}|\omega|^{-N}\leq \frac{1}{2C_N}.$$

Second, we construct a sequence of operators $H_{L_j}$ which satisfies assumption {\bf (SC)}. 

For fixed $L_j$, let us define the sequence of auxiliary operators
\beq
H'_{L_j}=-\Delta+a\sum_{\omega\in\Omega_{L_j}}V_{r_j}(\cdot-\omega)
\eeq
where $a$ denotes a real coupling parameter. Our goal is to choose $a=a_j$ for each $j$ in such a way that {\bf (SC)} is satisfied.

Fix $E\in(E_1,E_2)\setminus\Q$. Then the sequence $En^2 j^2$ equidistributes $\mod 1$. Hence, there exists a subsequence $(L_j)$ s. t. $EL_j^2$ remains bounded away from its neighbouring integers, as $j\to+\infty$.

By the min-max theorem we have that the eigenvalues of $H_{L_j}$ vary a continuous non-decreasing function of the real coupling parameter $a$. Denote by $\mu_j$ the smallest Laplace eigenvalue on the torus $\T^d_{L_j}$ which is greater or equal than $E$. Since each eigenvalue of $H_{L_j}$ tends to $-\infty$, as $a\to-\infty$, and it varies continuously as a function of $a$, there exists $a=a_j$ s. t. $E$ is in the spectrum. We denote this operator $H_{L_j}$ and the corresponding eigenvalue $\lambda_j$. Hence the sequence of operators $H_{L_j}$ and corresponding sequence of eigenvalues satisfies ${\bf (SC)}$.

\section{Limit theorems}

Assume $d\geq3$.
Let us consider the same sequence of operators 
\beq\label{gen}
H_L=-\Delta+ \sum_{\omega\in\Omega_L}\alpha_{\omega,L}V_r(\cdot-\omega), \quad a_{\omega,L}\in\C,
\eeq
on the tori $\T^d_L=\R^d/2\pi L\Z^d$, as $L\to+\infty$, as defined in the introduction.

We consider a  strictly increasing sequence of length scales $(L_j)_{j=1}^{+\infty}$, $L_j\nearrow+\infty$, and a sequence of eigenfunctions $u_j\in L^p(\T^d_{L_j})$, $\|u_j\|_{L^p(\T^d_{L})}=1$, $p>d/(d-2)$, of the operators $H_{L_j}$, $H_{L_j} u_j=\lambda_j u_j$, such that the eigenvalues $\lambda_j$ converge to a limit $E>0$, as $L_j\to+\infty$. A natural question concerns the existence of a nontrivial limit of this sequence in $L^1(\R^d)$. This question is related to the localization properties of the eigenfunctions $u_j$ in the thermodynamic limit and depends in a highly nontrivial way on the location of the scatterers $\omega\in\Omega_L$ and the coupling constants $a_{\omega,L}$. 

If the eigenfunctions are sufficiently delocalized in the thermodynamic limit, then, due to the $L^p$-normalization along the sequence of tori $\T^d_L$, the limit is $u=0$ on $\R^d$. On the contrary, if the eigenfunctions are sufficiently localized, due to the presence of disorder in the locations of the scatterers and/or the coupling constants, then non-trivial limits $u\in L^1(\R^d)$ may exist. If the scatterer locations and/or coupling constants are randomized in a suitable way, then one may observe exponential localization of the eigenfunctions almost surely, a phenomenon known as spectral Anderson localization that has been observed in a wide variety of models (e.g. \cite{BoKe05,DaSi02,GeKl13,KLNS12}). At the transition point between the localization and delocalization one expects the eigenfunctions to satisfy multifractal scaling laws.


\subsection{Notation and statement of results}
Fix $m\in\N$. Let us consider a partial differential operator
$$H=\sum_{|\alpha|\leq m} c_\alpha D^\alpha$$
where $\alpha=(\alpha_1,\cdots,\alpha_d)\in\N^d$ is a multi-index, $D^\alpha=\partial_{x_1}^{\alpha_1}\cdots\partial_{x_d}^{\alpha_d}$, $|\alpha|:=\alpha_1+\cdots+\alpha_d$, and $c_\alpha$ denote complex constant coefficients.

We will study solutions of the eigenvalue problem
\beq\label{eigen}
H_L u=\lambda u, \quad H_L:=H+V_L,
\eeq
on $\T^d_L$ in the limit as $L\to+\infty$, where
$$V_L=\sum_{\omega\in\Omega_L}\alpha_{\omega,L} V_r(\cdot-\omega).$$
In particular, we will consider the thermodynamic limit, where $N=N_L$ grows with the length parameter $L$ in such a way that $\lim_{L\to+\infty}N/L^d=\rho>0$. 

Let us consider an increasing sequence $(L_j)_{j=1}^{+\infty}$, $L_j\nearrow+\infty$, and assume that 
\beq\label{min_dist}
\liminf_{L_j\to+\infty}\delta_{L_j}>0,
\eeq
is satisfied along this sequence, where
$$\delta_L:=\inf\{d(x_k,x_l) \mid x_k, x_l\in \Omega_L, x_k\neq x_l\}$$
and $d(\cdot,\cdot)$ is the distance on $\T^d_L$.

We consider a sequence $u_j\in L^1(\T^d_{L_j})$ of solutions to \eqref{eigen} with $L=L_j$. We denote the corresponding sequence of eigenvalues by $(\lambda_j)$ and assume $\lim_{j\to+\infty}\lambda_j=E$ for some fixed $E\in[E_1,E_2]$.
 

\begin{defn}
Let $u\in L^1(\R^d)$.
With the above convention, we say that a sequence of eigenfunctions $u_j\in L^1(\T^d_{L_j})$ converges to $u$ {\em on the Fourier side} if for any $k\in\R^d$ and any sequence $\xi_j\in\Z^d$, s. t. $\xi_j/L_j\to k\in\R^d$, as $j\to+\infty$, we have
\beq\label{Fourier-conv}
\hat{u}_j\left(\frac{\xi_j}{L_j}\right)\to \hat{u}(k):=\int_{\R^d}u(x)e^{-ik\cdot x}dx.
\eeq
\end{defn}

We have the following result.
\begin{thm}\label{thermo_thm}
Let $d>m$.  Fix $p>d/(d-m)$ and take $p'=p/(p-1)$. Take $r=r_{L_j}=o(L_j^{-d/(d-mp')})$. Assume that \eqref{min_dist} is satisfied.
Let $(u_j)$ be a sequence of solutions to \eqref{eigen} with $L=L_j$, $\|u_j\|_{L^p(\T^d_{L_j})}=1$. Suppose that $(u_j)$ converges to $u\in L^1(\R^d)$ on the Fourier side in the sense of \eqref{Fourier-conv}. Then $u=0$.
\end{thm}

\subsection{Application to Schr\"odinger operators}
In particular, we may apply this result to any sequence of Schr\"odinger operators on $\T^d_{L_j}$, $d\geq3$,
$$H_j=-\Delta+\alpha_{L_j}\sum_{\omega\in\Omega_L}V_r(\cdot-\omega),$$ where $\Omega_L\subset\T^d_{L}$ is obtained from the restriction $\Omega\cap B_L$, $B_L=\pi[-L,L]^d$, of a point set in $\Omega\subset \R^d$, where we take $B_L$ as a fundamental domain for $\T^d_L$.

If $\Omega$ has sufficiently small disorder in the sense that there exists a sequence of length scales $L_j$, $L_j\to+\infty$, s. t. $\liminf_{j\to+\infty}\dist(\partial B_{L_j},\Omega)>0$, then condition \eqref{min_dist} is satisfied by the sequence of point sets $\Omega_{L_j}$ on the tori $\T^d_{L_j}$. 

Note that there exist {\em different coupling regimes} which give rise to different types of limit functions (or distributions). In particular, if $r\to 0$, as $L\to+\infty$, such operators may model the case of a scatterer with fixed  coupling $\alpha$ and $r$ which is fixed but much smaller than the wavelength. However, this requires a renormalization of $\alpha_L$, as $L\to+\infty$, as we will explain in section \ref{sec-coupling}. This type of renormalization is closely related to a semiclassical renormalization that has been studied in the context of \u{S}eba billiards in dimension $2$ \cite{Sh94,RU14} and is related to the singularity of the resolvent kernel of the Laplacian (note that in our case, the energy window is fixed and we take volume to infinity). This renormalization regime gives rise to $L^\infty$ functions in the thermodynamic limit which is the subject of the following corollary.\\

{\bf $L^\infty$ limits: }
We have the following corollary of Theorem \ref{thermo_thm} which shows that if a sequence of eigenfunctions converges to a limit in $L^\infty(\R^d)$ in the sense of \eqref{Fourier-conv}, then the amplitude of this function decays sufficiently slowly such that it is not integrable on $\R^d$.
\begin{coro}
Consider a sequence of eigenfunctions $v_j$ of $H_j$, $\|v_j\|_{L^\infty(\T^d_{L_j})}=1$, which converges in the sense of \eqref{Fourier-conv} to $v\in L^\infty(\R^d)$, $v\neq0$. Moreover, assume that $\|v_j\|_{L^p(\T^d_{L_j})}\to\ell_p>0$, as $j\to+\infty$ for some $p>d/(d-2)$. Then we have $v\notin L^1(\R^d)$.
\end{coro}
\begin{proof}
Assume for a contradiction that $v\in L^1(\R^d)$.
Define $u_j:=v_j/\|v_j\|_{L^p(\T^d_{L_j})}$ for $p>d/(d-2)$. We have that $u_j$ converges to $u$ on the Fourier side in the sense of \eqref{Fourier-conv} and thus we obtain $\hat{u}=\ell_p \hat{v}$. Our assumption implies that $u\in L^1(\R^d)$. We may now apply Theorem \ref{thermo_thm} to conclude $u=0$. However, this implies $v=\ell_p u=0$ which is a contradiction.
\end{proof}

{\bf General limits:}
If no suitable renormalization is carried out, then singular functions or even distributions may arise in the limit. The following theorem, which in fact implies Theorem \ref{thermo_thm}, gives a general characterization of limits on the Fourier side.
\begin{thm}\label{thermo_supp}
Let $d>m$. Fix $p>d/(d-m)$ and choose $p'=p/(p-1)$. Let $r=r_{L_j}=o(L_j^{-d/(d-mp')})$. Assume that \eqref{min_dist} is satisfied. Let $(u_j)$ be a sequence of eigenfunctions on $\T^d_{L_j}$ such that $\|u_j\|_{L^p(\T^d_{L_j})}=1$ and $\lambda_j\to E$, as $j\to+\infty$.
Moreover, let us suppose that for a sequence of lattice points $(\xi_j)$ in $\Z^d$ we have $\xi_j/L_j\to k$ for some $k\in\R^d$, as $j\to+\infty$, 
and that the sequence  $\widehat{u_j}(\xi_j/L_j)$ converges to a limit which we denote by
\beq
F_k:=\lim_{j\to+\infty}\widehat{u_j}\left(\frac{\xi_j}{L_j}\right).
\eeq
Then $F_k=0$ unless $P(k)=E$, where $P(\xi):=\sum_{|\alpha|\leq m}c_\alpha (i\xi)^{\alpha}$ denotes the symbol of $H$.
\end{thm}

\section{Motivation of coupling regimes and renormalization}\label{sec-coupling}

In this section we will study different regimes in terms of the growth or decay of the parameter $\alpha_L$.
Moreover, we will discuss which regimes model generic potentials whose support is fixed but whose diameter is small compared with the wavelength of the eigenfunction. For simplicity we will only deal with the case of real-valued potentials and will thus assume $\alpha_L\in\R$ throughout this section.

We recall that an eigenfunction $v_j$ is a solution of $H_j v_j=\lambda_j v_j$ which we may rewrite as
\beq
(-\Delta-\lambda_j)v_j=-\alpha_{L_j}V_{L_j} v_j, \quad V_{L_j}(x)=\sum_{\omega\in\Omega_{L_j}}V_r(x-\omega)
\eeq
which we Fourier transform as
\beq
(\frac{|\xi|^2}{L_j^2}-\lambda_j)\hat{v}_j(\frac{\xi}{L})=-\alpha_{L_j}\widehat{V_{L_j} v_j }(\frac{\xi}{L_j})
\eeq
where
$$\widehat{V_{L_j} v_j }(\frac{\xi}{L_j})=\sum_{\omega\in\Omega_L} \int_{\T^d_{L_j}}V_r(x-\omega)v_j(x)e^{-i\xi\cdot x/L_j}dx.$$

By summing over the lattice, we obtain
\beq
v_j(x)=-\alpha_{L_j}L_j^{2-d}\sum_{\xi\in\Z^d}\frac{\widehat{V_{L_j} v_j }(\xi/L_j)}{|\xi|^2-\lambda_j L_j^2}e^{i \xi\cdot x/L_j}.
\eeq

In order to understand the different coupling regimes let us examine the situation with a single scatterer place at a fixed point $\omega_0\in\T^d_{L_j}$. By translation invariance of the torus, and because there is a single scatterer, we may choose fundamental domains for the tori $\T^d_{L_j}$ in such a way that $\omega_0=0$ for all $j$. In this case we have
$$
v_j(x)=-\alpha_{L_j}L_j^{2-d}\sum_{\xi\in\Z^d}\frac{\widehat{V_r v_j }(\xi/L_j)}{|\xi|^2-\lambda_j L_j^2}e^{i \xi\cdot x/L_j}.
$$

Now if we assume that the potential is point-like, i. e. the diameter of its support is much smaller than the wavelength: $r\ll 1/\sqrt{\lambda_j}\asymp1/\sqrt{E}$, then the integrals in the summation are well-approximated by $v_j(\omega_0)e^{-i\xi\cdot \omega_0/L_j}\widehat{V_r}(\xi/L_j)$ so that we get the approximation (recall $\omega_0=0$)
\beq
v_j(x)\approx -\alpha_{L_j}L_j^{2-d}v_j(0)\sum_{\xi\in\Z^d}\frac{\widehat{V_r}(\xi/L_j)}{|\xi|^2-\lambda_j L_j^2}e^{i \xi\cdot x/L_j}
\eeq

If we now set $x=0$, and assume $v_j(0)\neq0$, then we get the approximate relation
\beq\label{sing_coupling}
1\approx -\alpha_{L_j}L_j^{2-d}\sum_{\xi\in\Z^d}\frac{\widehat{V_r}(\xi/L_j)}{|\xi|^2-\lambda_j L_j^2}.
\eeq

Note that we have the bound $$\Big|\widehat{V_r}\left(\frac{\xi}{L}\right)\Big|\lesssim_{V,N,d} \left(\frac{L}{r|\xi|}\right)^N,$$
where we recall the definition of $V_r$ by scaling the original potential $\tilde{V}$ by the small parameter $r$ and then periodizing. Therefore, the tail of the summation in \eqref{sing_coupling}, when $|\xi|\gg L/r$ is neglible and the sum is dominated by the terms $|\xi|\ll L/r$.

For the terms satisfying $|\xi|\ll L_j/r$ the wavelength of the complex exponential in the Fourier transform is much larger than the diameter of support of the potential $V_r$: $r\ll L_j/|\xi|$. In this regime we have $\widehat{V_r}(\xi/L_j)\approx \|V_r\|_{L^1(\T^d_{L_j})}=\|\tilde{V}\|_{L^1(\R^d)}$. Therefore, the sum on the r.h.s. of \eqref{sing_coupling} is dominated by a finite series
\beq
1\approx -\alpha_{L_j}L_j^{2-d}\|\tilde{V}\|_{L^1(\R^d)}\sum_{|\xi|^2\leq c L^2/r^2}\frac{1}{|\xi|^2-\lambda_j L_j^2}
\eeq

The sum on the r.h.s. can be written as
$$\sum_{n\leq cL^2/r^2}\frac{r_d(n)}{n-\lambda_j L^2}, \quad r_d(n):=\#\{\xi\in\Z^d \mid |\xi|^2=n\}.$$
It is known that $r_d(n)\asymp n^{d/2-1}$ for $d\geq 5$ (cf. Thm. 20.2, in \cite{IwKo04}, p. 456), and $r_d(n)=O(n^{d/2-1+o(1)})$ for $d=3,4$.

Let us denote by $n_j^-$ the largest integer s. t. $r_d(n_j)\neq0$ and $n_j^-<\lambda_j L_j^2$, and by $n_j^+$ the smallest integer s. t. $r_d(n_j)\neq0$ and $n_j^+>\lambda_j L_j^2$.

We divide the summation up as follows
\beq\label{key_expansion}
\sum_{n\leq cL^2/r^2}\frac{r_d(n)}{n-\lambda_j L^2}=\frac{r_d(n_j^+)}{\Delta_+}-\frac{r_d(n_j^-)}{\Delta_-}+S_j
\eeq
where
\beq
S_j=\sum_{n<n_j^-}\frac{r_d(n)}{n-\lambda_j L^2}+\sum_{n_j^+<n\leq cL_j^2/r^2}\frac{r_d(n)}{n-\lambda_j L_j^2}
\eeq
and $\Delta_\pm=|n_j^\pm-\lambda_j L_j^2|$. 

We will now use identity \eqref{key_expansion} to determine the coupling regime for $\alpha_L$, where the potential with rapidly shrinking support, $r_L\to0$, models the potential with fixed support $r=r_0$ and fixed coupling $\alpha_L=\alpha_0$. In order to do this we note that we have to compute the asymptotic of the two sums, and in particular their dependence on the parameters $L$ and $r$. 

We have the following asymptotic which is proved in section \ref{sec-asymp}:
\beq
S_j \sim c_d\left(\frac{L_j}{r}\right)^{d-2}(1+O(r/L_j))+O(L_j^{d-1}).
\eeq

If we substitute these asymptotics in \eqref{key_expansion} and recall that $n_j^\pm \asymp \lambda_j L_j^2$, then we obtain the relation
\beq
-\frac{1}{\alpha_L}\approx \|\tilde{V}\|_{L^1(\R^d)}(E^{d/2-1}\left\{\frac{1}{\Delta_+}-\frac{1}{\Delta_-}\right\}+c_d r_L^{-d+2})
\eeq
For simplicity, let us set $\|\tilde{V}\|_{L^1(\R^d)}, E=1$, and rewrite this relation as
\beq\label{sing_comp}
\frac{1}{\alpha_L}+r_L^{-d+2}\approx \frac{1}{\Delta_-}-\frac{1}{\Delta_+}
\eeq

From this relation we may distinguish different coupling regimes:
\begin{itemize}
\item[(i)] {\bf Fixed coupling, fixed support: $\alpha_L=\alpha_0$, $r_L=r_0$.}\\ 
In this regime, the distances $\Delta_+$ and $\Delta_-$ are of comparable size.
\\
\item[(ii)] {\bf Small coupling, fixed support: $\alpha_L\to 0$, $r_L=r_0$.}\\ 
If we take $\alpha_L\to 0$, as $L\to+\infty$, but the diameter of the support of the potential $r_L=r_0$ remains fixed, then the blow-up of $1/\alpha_L$ on the l.h.s. of \eqref{sing_comp} forces a singularity on the r.h.s., as $L\to+\infty$: if $\alpha_L\to0^+$, then $\Delta_-\to 0$ and $\Delta_-\ll \Delta_+$. On the other hand, if $\alpha\to0^-$, then $\Delta_+\to 0$ and $\Delta_-\gg \Delta_+$.
\\
\item[(iii)]{{\bf Small coupling, small support: $\alpha_L\to 0$, $r_L\to 0$.}\\
In this regime, since we are taking $r_L\to0$, $\alpha_L$ does not correspond to a physical coupling parameter as in regimes (i) and (ii). Rather the relation \eqref{sing_comp} describes a renormalization of the coupling parameter. For given $\alpha_0$, $r_0$ one may model a scatterer as described in regime (i) by renormalizing $\alpha_L$ in such a way that the quantity $\Delta_-^{-1}-\Delta_+^{-1}$ matches in both regimes which gives rise to the following relation which describes the renormalization of $\alpha_L$ in terms of $r_L$, $\alpha_0$ and $r_0$:
\beq\label{renorm}
\frac{1}{\alpha_L}=-r_L^{-d+2}+\frac{1}{\alpha_0}+r_0^{-d+2}
\eeq
Hence $1/\alpha_L$ features a singularity, as $L\to+\infty$, and the information about physical coupling and size of the scatterer are encoded in the lower order terms.}
\end{itemize}

\section{Proof of Theorems \ref{thermo_thm} and \ref{thermo_supp}}

\subsection{Proof of Theorem \ref{thermo_supp}}
We introduce a smooth cutoff $\tilde{\chi}\in C^\infty_c(\R^d)$ s. t. $\tilde{\chi}=1$ on $B(0,1)$, $0\leq\tilde\chi\leq 1$ on $B(0,2)\setminus B(0,1)$ and $\tilde{\chi}=0$ outside $B(0,2)$. We define the following cutoff on $\T^d_L$:
$$\chi_r(x)=\sum_{\xi\in2\pi\Z^d}\tilde{\chi}\left(\frac{x-L\xi}{r}\right)$$
Moreover, for the set of points $\Omega_L:=\{x_1,\cdots,x_N\}\subset\T^d_L$ we define
$$\chi_L(x)=\sum_{k=1}^N \chi_r(x-x_k).$$
By construction we have $\supp \chi_L=\bigcup_{k=1}^N B(x_k,2r)$. Moreover, if $4r<\delta_L$, then all balls are disjoint and $0\leq\chi_{L}\leq 1$. In particular, we note that $\chi_L(x)=1$ if $x\in \bigcup_{k=1}^N B(x_k,r)$. We recall from the definition of the potential $V_L$ that $\supp V_L\subset\bigcup_{k=1}^N B(x_k,r)$. Hence, we see that $\supp(1-\chi_L)$ does not intersect the balls $B(x_k,r)$, $k=1,\cdots,N$, and therefore $(1-\chi_L)V_L=0$, because $\supp (1-\chi_L)\cap\supp V_L=\emptyset$.

Now let us define the following quasi-mode on $\T^d_{L_j}$:
$$
\Psi_{L_j}:=(1-\chi_{L_j})u_j.
$$

\subsubsection{Motivation: Schr\"odinger operators}

Take $H=-\Delta$. 

So we have $(\Delta+\lambda_j)u_j =V_{L_j}u_j$.
Moreover, 
$$
(\Delta+\lambda_j)\Psi_{L_j}=2\nabla\chi_{L_j}\cdot \nabla u_j+(\Delta\chi_{L_j}) u_j
$$
where we used $(1-\chi_{L_j})(\Delta+\lambda_j)u_j =(1-\chi_{L_j})V_{L_j}u_j=0$, and we recall that by the construction of the cutoff $\chi_{L_j}$ we have $\supp (1-\chi_{L_j})\cap\supp V_{L_j}=\emptyset$.

If we apply the Fourier transform on $\T^d_L$ to the identity above, we obtain
\begin{equation}
\begin{split}
\left(\lambda_j-\frac{|\xi^2|}{L^2}\right)\widehat{\Psi_{L_j}}\left(\frac{\xi}{L}\right) = & \,2\int_{\T^d_L}\nabla\chi_{L_j}(x)\cdot \nabla u_j(x)e^{-i \xi\cdot x/L}dx\\
&+\int_{\T^d_L}(\Delta\chi_{L_j}(x)) u_j(x) e^{-i \xi\cdot x/L} dx
\end{split}
\end{equation}
We estimate the first integral on the r.h.s. by
\begin{equation*}
\begin{split}
\Big|\int_{\T^d_L}\nabla\chi_{L_j}(x)\cdot \nabla u_j(x)e^{-i \xi\cdot x/L}dx\Big|
=&\Big| \int_{\T^d_L}(\Delta\chi_{L_j}(x)-i\nabla\chi_{L_j}(x)\cdot \frac{\xi}{L})u_j(x)e^{-i\xi\cdot x/L}dx\Big|\\
=&\Big\| \Delta\chi_{L_j}(x)-i\nabla\chi_{L_j}(x)\cdot \frac{\xi}{L}\Big\|_{L^{p'}(\T^d_L)} \|u_j\|_{L^p(\T^d_L)}\\
\leq& (\|\Delta\chi_{L_j}\|_{L^{p'}(\T^d_L)}+\frac{|\xi|}{L}\||\nabla\chi_{L_j}|\|_{L^{p'}(\T^d_L)})\|u_j\|_{L^p(\T^d_L)}
\end{split}
\end{equation*}
where $1/p+1/p'=1$.

This yields the estimate
$$
\Big|\frac{|\xi|^2}{L^2}-\lambda_j\Big| \Big|\widehat{\Psi_{L_j}}\left(\frac{\xi}{L}\right)\Big| \leq (3\|\Delta\chi_{L_j}\|_{L^{p'}(\T^d_L)}+2\frac{|\xi|}{L}\||\nabla\chi_{L_j}|\|_{L^{p'}(\T^d_L)})\|u_j\|_{L^p(\T^d_L)}
$$

We have 
$$\|\Delta\chi_{L_j}\|_{L^{p'}(\T^d_L)}\leq c_1 r^{-2}(\Vol\supp \chi_{L_j})^{1/p'}$$
and $$\||\nabla\chi_{L_j}|\|_{L^{p'}(\T^d_L)}\leq c_2 r^{-1} (\Vol\supp \chi_{L_j})^{1/p'}$$
where the constants $c_1$, $c_2$ only depend on derivatives of the fixed cutoff $\chi$.

We have $$\Vol\supp\chi_{L_j}\leq \sum_{k=1}^N \Vol (B(x_k,r))=c_d Nr^d,$$
where $c_d$ denotes the volume of the $d$-dimensional unit ball.

Thus, we have
\beq\label{key_est_schr}
\Big|\frac{|\xi_j|^2}{L_j^2}-\lambda_j\Big| \Big|\widehat{\Psi_{L_j}}\left(\frac{\xi_j}{L_j}\right)\Big|\leq C(\chi,d) N^{1/p'}r^{d/p'-1}(r^{-1}+\frac{|\xi_j|}{L_j})\|u_j\|_{L^p(\T^d_{L_j})}
\eeq

Recall our assumption $q>d/(d-2)$, which implies $p'=p/(p-1)<d/2.$ If we recall that $\frac{\xi_j}{L_j}\to k\in\R^d$, $\|u_j\|_{L^p(\T^d_{L_j})}=1$,
$N \asymp \rho L_j^d$, then, for the choice $r=r_L=o(L^{-d/(d-2p')})$, the r.h.s. of \eqref{key_est_schr} tends to $0$. In particular, we may conclude that $\widehat{\Psi_{L_j}}\left(\frac{\xi_j}{L_j}\right)\to 0$, as $j\to+\infty$, unless $|k|^2=E$.

The triangle inequality yields the estimate
$$|\widehat{u_j}\left(\frac{\xi_j}{L_j}\right)|\leq\Big|\widehat{u_j}\left(\frac{\xi_j}{L_j}\right)-\widehat{\Psi_{L_j}}\left(\frac{\xi_j}{L_j}\right)\Big| + \Big|\widehat{\Psi_{L_j}}\left(\frac{\xi_j}{L_j}\right)\Big|.$$

Moreover, 
$$\Big|\widehat{u_j}\left(\frac{\xi_j}{L_j}\right)-\widehat{\Psi_{L_j}}\left(\frac{\xi_j}{L_j}\right)\Big|\leq \int_{\T^d_L}\chi_{L_j}(x)|u_j(x)| dx
\leq c_3 (\Vol\supp\chi_{L_j})^{1/p'}\|u_j\|_{L^p(\T^d_{L_j})}$$
which tends to $0$ for our choice $r_{L_j}$.

We conclude that   $F_k=\lim_{j\to+\infty}\widehat{u_j}\left(\frac{\xi_j}{L_j}\right)=0$ vanishes unless $k\in\IS^{d-1}_{\sqrt{E}}$.

\subsubsection{General case}

We now give the proof for a general constant coefficient partial differential operator $H$. 

We have
$$
(H-\lambda_j)\Psi_{L_j}=\sum_{|\alpha|\leq m} c_\alpha \sum_{\substack{\alpha=\beta+\beta' \\ |\beta|>0}}C_\beta D^{\beta}\chi_{L_j} \cdot D^{\beta'}u_j
$$
where we used the product identity
$$D^\alpha(fg)=\sum_{\alpha=\beta+\beta'}C^\alpha_\beta D^\beta f \cdot D^{\beta'}g, \quad C^\alpha_\beta=\prod_{j=1}^d {{\alpha_j}\choose{\beta_j}},$$
and the sum runs over multi-indices $\beta,\beta'$.

If we apply the Fourier transform on $\T^d_L$ to the identity above, we obtain
\begin{equation}
\begin{split}
\left(P\left(\frac{\xi}{L}\right)-\lambda_j\right)&\widehat{\Psi_{L_j}}\left(\frac{\xi}{L}\right) \\
= & \sum_{|\alpha|\leq m} c_\alpha \sum_{\substack{\alpha=\beta+\beta' \\ |\beta|>0}}C^\alpha_{\beta}\widehat{(D^{\beta}\chi_{L_j} \cdot D^{\beta'}u_j)}\left(\frac{\xi}{L}\right)
\end{split}
\end{equation}

We estimate 
\begin{equation*}
\begin{split}
\Big|\widehat{(D^{\beta}\chi_{L_j} \cdot D^{\beta'}u_j)}\left(\frac{\xi}{L}\right)\Big|
=&\Big| \int_{\T^d_L}(D^{\beta}\chi_{L_j}(x) \cdot D^{\beta'}u_j(x))e^{-i\xi\cdot x/L}dx\Big|\\
=&\Big| \int_{\T^d_L}D^{\beta'}((D^{\beta}\chi_{L_j}(x) e^{-i\xi\cdot x/L}) u_j(x))dx\Big|\\
=&\Big\| D^{\beta'}((D^{\beta}\chi_{L_j}(x) e^{-i\xi\cdot x/L})\Big\|_{L^{p'}(\T^d_L)} \|u_j\|_{L^p(\T^d_L)}
\end{split}
\end{equation*}
where $1/p'+1/p=1$.

We compute
\beq
\begin{split}
D^{\beta'}((D^{\beta}\chi_{L_j}(x) e^{-i\xi\cdot x/L})
=\sum_{\beta'=\gamma+\gamma'}C^{\beta'}_{\gamma}(D^{\beta+\gamma}\chi_{L_j})(x) \left(\frac{-i\xi}{L}\right)^{|\gamma'|}e^{-i\xi\cdot x/L}
\end{split}
\eeq
and, thus,
\beq
\begin{split}
\Big\| D^{\beta'}((D^{\beta}\chi_{L_j}(x) e^{-i\xi\cdot x/L})\Big\|_{L^{p'}(\T^d_L)}&
\leq \sum_{\beta'=\gamma+\gamma'}|C^{\beta'}_{\gamma}|\|D^{\beta+\gamma}\chi_{L_j}\|_{L^{p'}(\T^d_L)} \left(\frac{|\xi|}{L}\right)^{|\gamma'|}
\end{split}
\eeq

Now, $$\|D^{\beta+\gamma}\chi_{L_j}\|_{L^{p'}(\T^d_L)}\leq \|D^{\beta+\gamma}\chi_{L_j}\|_{L^{\infty}(\T^d_L)}(\Vol\supp\chi_{L_j})^{1/p'}$$
and $$\|D^{\beta+\gamma}\chi_{L_j}\|_{L^{\infty}(\T^d_L)}\lesssim_\chi r^{-|\beta+\gamma|}\leq r^{-m}$$
which yields
$$\Big\| D^{\beta'}((D^{\beta}\chi_{L_j}(x) e^{-i\xi\cdot x/L})\Big\|_{L^{p'}(\T^d_L)} \lesssim_{m,\chi} r^{-m}(\Vol\supp \chi_{L_j})^{1/p'}\left(\frac{|\xi|}{L}\right)^m$$
where the constant depends only on derivatives of the fixed cutoff $\chi$.

We deduce the estimate
\beq
\begin{split}
\Big|P\left(\frac{\xi}{L}\right)-\lambda_j\Big|& \Big|\widehat{\Psi_{L_j}}\left(\frac{\xi}{L}\right)\Big| \\
\leq \sum_{|\alpha|\leq m} |c_\alpha| & \sum_{\substack{\alpha=\beta+\beta' \\ \beta>0}}|C^\alpha_{\beta}|
 \Big\| D^{\beta'}((D^{\beta}\chi_{L_j})e_{-\xi/L})\Big\|_{L^{p'}(\T^d_L)} \|u_j\|_{L^p(\T^d_L)}\\
 &\lesssim _{m,\chi} r^{-m}(\Vol\supp \chi_{L_j})^{1/p'} \left(\frac{|\xi|}{L}\right)^m \|u_j\|_{L^p(\T^d_L)}
\end{split}
\eeq

We have $$\Vol\supp\chi_{L_j}\leq \sum_{k=1}^N \Vol (B(x_k,r))=c_d Nr^d,$$
where $c_d$ denotes the volume of the $d$-dimensional unit ball.

Thus, we have
\beq\label{key_est}
\Big|P\left(\frac{\xi_j}{L_j}\right)-\lambda_j\Big| \Big|\widehat{\Psi_{L_j}}\left(\frac{\xi_j}{L_j}\right)\Big|\lesssim_{m,\chi,d} N^{1/p'}r^{d/p'-m}\left(\frac{|\xi_j|}{L_j}\right)^m\|u_j\|_{L^p(\T^d_{L_j})}
\eeq

Recall our assumptions $\xi_j/L_j\to k\in\R^d$, $\|u_j\|_{L^p(\T^d_{L_j})}=1$,
$N \asymp \rho L_j^d$, as well as $p>d/(d-m)$, which implies $p'=p/(p-1)<d/m$. 

Then, for the choice $r=r_L=o(L^{-d/(d-mp')})$, the r.h.s. of \eqref{key_est} tends to $0$. In particular, we may conclude that $\widehat{\Psi_{L_j}}\left(\frac{\xi_j}{L_j}\right)\to 0$, as $j\to+\infty$, unless $P(k)=E$.

The triangle inequality yields the estimate
$$|\widehat{u_j}\left(\frac{\xi_j}{L_j}\right)|\leq\Big|\widehat{u_j}\left(\frac{\xi_j}{L_j}\right)-\widehat{\Psi_{L_j}}\left(\frac{\xi_j}{L_j}\right)\Big| + \Big|\widehat{\Psi_{L_j}}\left(\frac{\xi_j}{L_j}\right)\Big|.$$

Moreover, 
$$\Big|\widehat{u_j}\left(\frac{\xi_j}{L_j}\right)-\widehat{\Psi_{L_j}}\left(\frac{\xi_j}{L_j}\right)\Big|\leq \int_{\T^d_L}\chi_{L_j}(x)|u_j(x)| dx
\leq (\Vol\supp\chi_{L_j})^{1/p'}\|u_j\|_{L^p(\T^d_{L_j})}$$
which tends to $0$ for our choice $r_{L_j}$.

We conclude that $F_k=\lim_{j\to+\infty}\widehat{u_j}\left(\frac{\xi_j}{L_j}\right)=0$ vanishes unless $k\in\{k'\in\R^d : P(k')=E\}$ which proves the claim.

\subsection{Proof of Theorem \ref{thermo_thm}}

We apply Lemma \ref{thermo_supp}. By assumption we have $\hat{u}(k)=0$ if $P(k)\neq E$. Since $u\in L^1(\R^d)$ implies $\hat{u}\in L^\infty(\R^d)$,
it follows by Fourier inversion
$$
|u| \lesssim_d \|\hat{u}\|_{L^1(\R^d)}\leq \Vol(\supp{\hat{u}})\|\hat{u}\|_{L^{\infty}(\R^d)}=0,
$$
because $\Vol\supp \hat{u}\leq \Vol(\{k : P(k)=E\})=0$, since $P$ is a non-zero polynomial.

\section{Asymptotics for lattice sums}\label{sec-asymp}

We may evaluate the lattice sums using summation by parts 
\beq
\sum_{n_A<n\leq n_B}r_d(n)f(n) = N(n_B)-N(n_{A+1})-\int_{n_{A+1}}^{n_B}f'(t)N(t)dt
\eeq
where $$N(t)=\sum_{n\leq t}r_d(n)=c_d t^{d/2}+O(t^{(d-1)/2})$$ and $f(t)=(t-\lambda_jL_j^2)^{-1}$.

Using integration by parts and the above asymptotic we get
\beq
\begin{split}
\sum_{n_A<n\leq n_B}r_d(n)f(n) = &\frac{d c_d}{2}\int_{n_{A+1}}^{n_B}f(t)t^{d/2-1}dt \\
&+O(f(n_B)n_B^{(d-1)/2})+O(f(n_{A+1})n_{A+1}^{(d-1)/2})\\
&+O\left(\int_{n_{A+1}}^{n_B}|f'(t)|t^{(d-1)/2}dt\right).
\end{split}
\eeq

For the first sum we have $n_A=0$ and $n_B\leq n_j^--1$. The two error terms arising from the boundary of the integral are bounded by $(\lambda_j L^2)^{(d-1)/2}$, because $|f(n_B)|=O(1)$ and the error term arising from $n_{A+1}$ is clearly of order $O(1)$. The integral error term is bounded by
\beq
\begin{split}
\int_1^{n_j^- -1}\frac{t^{(d-1)/2}dt}{(t-\mu)^2} &\leq \mu^{(d-3)/2}\int^{1-1/\mu}_{0}\frac{s^{(d-1)/2}ds}{(s-1)^2}\\
&\leq \mu^{(d-3)/2}\int^{1-1/\mu}_{0}\frac{ds}{(s-1)^2} \\
&\lesssim \mu^{(d-1)/2}= (\lambda_j L^2)^{(d-1)/2}
\end{split}
\eeq

The main integral term has asymptotic
\beq
\begin{split}
& \int_{1}^{n_B}\frac{t^{d/2-1}dt}{t-\mu}\\
&=\mu^{d/2-1}\int_{1/\mu}^{n_B/\mu}\frac{s^{d/2-1}ds}{s-1}\\
&\lesssim (\lambda_jL_j^2)^{d/2-1}\log(\lambda_j L_j^2)
\end{split}
\eeq
so that we have 
\beq
\sum_{n<n_L^-}\frac{r_d(n)}{n-\lambda_j L_j^2} \lesssim L_j^{d-1}.
\eeq

Next we estimate the error terms for the summation by parts of the second sum. The integral error is bounded by 
\beq
\begin{split}
\int_{n_j^++1}^{cL^2/r^2}\frac{t^{(d-1)/2}dt}{(t-\mu)^2}=&\mu^{(d-3)/2}\int^{cL^2/(r^2\mu)}_{1+\Omega(1/\mu)}\frac{s^{(d-1)/2}ds}{(s-1)^2}\\
=&\mu^{(d-3)/2}(\int_{1+\Omega(1/\mu)}^2 \frac{s^{(d-1)/2}ds}{(s-1)^2}+\int_2^{cL^2/(r^2\mu)} \frac{s^{(d-1)/2}ds}{(s-1)^2})\\
=&c_d \mu^{(d-3)/2}( (\frac{L^2}{r^2\mu})^{(d-3)/2}+O(\mu))\\
=&c_d\left(\frac{L}{r}\right)^{d-3}+O(L^{d-1})
\end{split}
\eeq
and the main integral gives
\beq
\begin{split}
 \int_{\lambda_j L_j^2+1}^{cL^2/r^2}\frac{t^{d/2-1}dt}{t-\lambda_j L_j^2}
&=(\lambda_jL_j^2)^{d/2-1}\int_{1+\lambda_j^{-1}L_j^{-2}}^{c\lambda_j^{-1}r^{-2}}\frac{s^{d/2-1}ds}{s-1}\\
&\sim (\lambda_jL_j^2)^{d/2-1}\{c_d(\lambda_j^{-1}r^{-2})^{d/2-1}-\log(\lambda_j L_j^2)\}\\
&= c_d\left(\frac{L_j}{r}\right)^{d-2}+O(L_j^{d-2}\log L_j).
\end{split}
\eeq

The boundary terms are estimated as follows (recall $n_A=n_j^+$)
$$f(n_{A+1})n_{A+1}^{(d-1)/2}\lesssim (\lambda_j L_j^2)^{(d-1)/2}\lesssim L_j^{d-1}$$
and (recall $n_B=cL^2/r^2$)
$$f(n_B)n_B^{(d-1)/2}\lesssim \left(c\frac{L_j^2}{r^2}\right)^{-1}\left(c\frac{L_j^2}{r^2}\right)^{(d-1)/2}\lesssim \left(\frac{L_j}{r}\right)^{d-3}.$$

This yields the following asymptotic for the second lattice sum
\beq
\sum_{n_L^+<n\leq cL^2/r^2}\frac{r_d(n)}{n-\lambda_j L_j^2} \sim c_d(\frac{L_j}{r})^{d-2}(1+O(r/L_j))+O(L_j^{d-1}).
\eeq

\end{document}